\documentclass[aps,prb,twocolumn,groupedaddress,amsmath,showpacs]{revtex4-1}
\usepackage{xcolor}
\usepackage{bbold}
\usepackage{graphicx}
\usepackage{siunitx}
\usepackage[caption=false]{subfig}

\begin{document}

\title{Topological entanglement entropy of three-dimensional Kitaev
  model}

\author{N. C. Randeep, Naveen Surendran}
\email[]{naveen.surendran@iist.ac.in}

\affiliation{Indian Institute of Space Science and Technology,
  Valiamala, Thiruvananthapuram-695547, Kerala, India}

\date{\today}

\begin{abstract}
We calculate the topological entanglement entropy (TEE) for a
three-dimensional hyperhoneycomb lattice generalization of Kitaev's
honeycomb lattice spin model. We find that for this model TEE is not
directly determined by the total quantum dimension of the system. This
is in contrast to general two dimensional systems and many three
dimensional models, where TEE is related to the total quantum
dimension. Our calculation also provides TEE for a three-dimensional
toric-code-type Hamiltonian that emerges as the effective low-energy
theory for the Kitaev model in a particular limit.
\end{abstract}


\maketitle

\section{Introduction}
Recent years have seen an explosion of research activity in the study
of new quantum phases of matter. Departing from the Landau paradigm of
classifying phases based on symmetries and local order parameters,
such phases, which are gapped, are immune to distinction through any
local operators. Instead, they are characterized by fractional
excitations and ground state degeneracy dependent on the topology of
the space.

Fractional excitations arise due to nontrivial long-range correlations
in the ground state. Bipartite entanglement entropy of a system
encapsulates such correlations by measuring the extent to which one
partition is entangled with the other. It is defined as the von
Neumann entropy of the reduced density matrix of one of the
partitions, which is obtained by taking partial trace with respect to
the degrees of freedom belonging to the other partition.

In gapped systems, the leading contribution to entanglement entropy
comes from a region around the boundary of the two partitions but
lying within the correlation length. As a result, the entanglement
entropy obeys an area law: it is proportional to the ``area'' of the
boundary between the two partitions\cite{Sre(93)}.

In a seminal work, Kitaev and Preskill\cite{KitPre(06)} and,
independently, Levin and Wen\cite{LevWen(06)}, showed that in
two-dimensional gapped systems the entanglement entropy $S$ contains,
apart from the term proportional to the length of the boundary $L$, a
constant term that depends only on the topology of the boundary
curve. They also showed that this constant is related to the total
quantum dimension of the system $\mathcal{D} = \sqrt{\sum_a d_a^2}$,
$d_a$ being the quantum dimension of $a$-type anyon. Specifically,
$S=\alpha L - b_0 \gamma$, where $\alpha$ is a positive non-universal
constant, $b_0$ is the zeroth Betti number (number of connected
components) of the boundary, and
\begin{align}
  \gamma &= \log \mathcal{D}.
  \label{gamma}
\end{align}

$\mathcal{D}$ is greater than one only when the system is
topologically ordered and has anyonic excitations. Thus, a nonzero
$\gamma$ is a signature of topological order and $\gamma$ is therefore
called the topological entanglement entropy (TEE).

Topological entanglement entropy has been calculated for two
dimensional models such as the toric code \cite{Kit(03),HamIon(05)}
and Kitaev's honeycomb lattice model \cite{Kit(06),YaoQi(10)},
verifying Eq. (\ref{gamma}).

A natural question then is: In higher dimensions $D$, in particular
for $D=3$, does a constant term in entanglement entropy imply
topological order?  Grover et al.\cite{GroTur(11)} have addressed
this question and, based on an expansion of local contributions to the
entropy in terms of curvature and its derivatives, they have found
that in three dimensions (and in general, for any odd $D$) a constant
term can arise in a generic gapped system purely from local
correlations. That is, a non zero $\gamma$ does not necessarily imply
topological order.

Furthermore, two-dimensional boundary surfaces have two topological
invariants---in addition to zeroth Betti number $b_0$, there is also
$b_1$, the first Betti number (number of noncontractible loops)---and
TEE, in general, can depend on both:
\begin{align}
  S &= \alpha A - b_0 \gamma_0 + \frac{b_1}{2} \gamma_1,
  \label{gamma-3d}
\end{align}
where $A$ is the area of the boundary and $\alpha,~\gamma_0$ and
$\gamma_1$ are constants. However, for compact surfaces $b_0$ and
$b_1$ are not independent and are related through the Euler
characteristic, $\chi = 2b_0 - b_1$, which can be thought of as a sum
of local terms and therefore be absorbed into the area term; thus
$\gamma_0$ and $\gamma_1$ are not independent topological
entropies\cite{GroTur(11)}.

Even though trivial phases in 3D may also give rise to a constant term
in the entropy, the topological contribution can still be extracted by
considering various carefully chosen partitioning of the system and
then taking an appropriate linear combination of the corresponding
entropies \cite{KitPre(06),LevWen(06),CasCha(08),GroTur(11)}. In the
process local, non-topological contributions are eliminated.

In three dimensions, TEE has been calculated for some exactly solvable
models. These include: the cubic lattice toric code\cite{CasCha(08)},
general quantum double models\cite{IblPer(09),IblPer(10),GroTur(11)},
and Walker-Wang models\cite{WalWan(12),KeyBur(13),BulPac(16)}. In all
these cases, $\gamma_0 = \ln \mathcal{D}$, which is similar to the
general case in two dimensions ($\mathcal{D}$ being the total quantum
dimension). TEE has also been calculated\cite{MonHug(14)} for
three-dimensional Ryu-Kitaev model\cite{Ryu(09)}, which is a
generalization of Kitaev's honeycomb lattice
model\cite{Kit(06)}. However, for this model it is not clear to us
what the total quantum dimension is and we have not been able to check
the above relation. It is then interesting to examine other
three-dimensional models and see whether such a relation between TEE
and $\mathcal{D}$ exists or not.

Partly motivated by the above question, in this paper we calculate TEE
of another three-dimensional generalization of Kitaev model defined on
the hyperhoneycomb lattice\cite{ManSur(09)}. We find that $\gamma_0 =
\ln 2$ and $\gamma_1=0$. For this model the total quantum dimension
$\mathcal{D}=\sqrt{2}$. Thus, our calculation
provides an example of a three-dimensional model for which $\gamma_0
\neq \ln \mathcal{D}$, unlike in the other 3D models mentioned above.

Kitaev model on hyperhoneycomb lattice has been of interest recently in
the context of certain iridium oxides\cite{KimAna(14)}. See
Ref. \onlinecite{ObrHer(16)} for a comprehensive study of the phases
of Kitaev model in three dimensions and Ref. \onlinecite{MatHer(17)}
for a study of its entanglement spectrum.

The rest of the paper is organized as follows. In Sec. \ref{s-km} we
briefly review Kitaev model on the hyperhoneycomb lattice and in
Sec. \ref{s-tee}, following the method of Yao and Qi\cite{YaoQi(10)},
we calculate its TEE. We conclude with a discussion in
Sec. \ref{s-conclusion}.

\section{\label{s-km} Kitaev model on hyperhoneycomb lattice}

Kitaev's honeycomb lattice spin model\cite{Kit(06)} has become a
paradigm system in the study of topological order in
quantum many-body systems. It is an exactly solvable spin-$1/2$ system
with two phases that respectively support Abelian and non-Abelian
anyons. Many proposals have been put forth for possible physical
realizations of Kitaev Hamiltonian (see Ref. \onlinecite{WinTsi(17)}
for a detailed review).

\subsection{Hamiltonian}

Kitaev's original model is defined on a honeycomb lattice with
spin-$1/2$ degrees of freedom at each site. Honeycomb lattice has
three types of links corresponding to three different orientations,
which are respectively labeled $x$-, $y$- and $z$-links. Neighboring
spins interact via Ising interaction, with the component of the Pauli
matrices in the interaction being same as the link-type. In
general, Kitaev Hamiltonian can be defined on any trivalent lattice in
which the links can be similarly labeled and in such a way that at
each site the three links are all of different type. Then the
Hamiltonian is
\begin{align}
H &=-J_{x}\sum_{x\mathrm{-links}} \sigma_j^x
\sigma_k^x - J_{y}\sum_{y\mathrm{-links}} \sigma_j^y
\sigma_k^y - J_{z}\sum_{z\mathrm{-links}} \sigma_j^z \sigma_k^z
\label{kmh}
\end{align}

\begin{figure}
    \begin{center}
      \includegraphics[scale=.5]{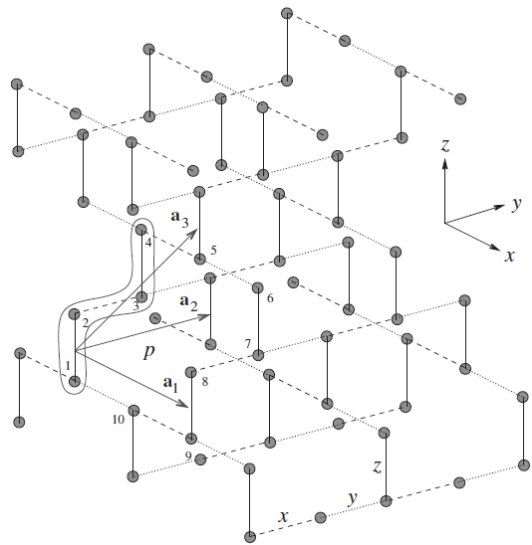}
    \caption{The $3$D lattice: the four sites labeled $(1-4)$
      constitute a unit cell and
      $\textbf{a}_{1}$,~$\textbf{a}_{2}$,~$\textbf{a}_{3}$ are the
      basis vectors. $x$-,~$y$- and $z$-links are represented by dashed,
      dotted and bold lines, respectively.}
    \label{f-lattice}
    \end{center}
    \end{figure}

In this paper we consider Kitaev Hamiltonian defined on the
three-dimensional lattice introduced in Ref. \onlinecite{ManSur(09)}
(see Fig. \ref{f-lattice}). The lattice we consider has the same
connectivity as the hyperhoneycomb lattice and is therefore
topologically equivalent to it. Kimchi et al.\cite{KimAna(14)} have
proposed a Kitaev-Heisenberg Hamiltonian---Kitaev model with
additional Heisenberg interactions---on the hyperhoneycomb lattice to
model certain iridium oxides. Their proposal is based on a mechanism
introduced by Jackeli and Khaliullin\cite{JacKha(09)}, by which the
bond-anisotropic Kitaev interaction can arise from strong spin-orbit
coupling.

The unit cell of the hyperhoneycomb lattice contains four sites, as
shown in Fig. \ref{f-lattice}. The basis vectors are given by ${\bf a}_1 =
2 \hat {\bf x},~{\bf a}_2 = 2 \hat {\bf y},~{\bf a}_3 =\hat {\bf x} +
\hat {\bf y} + 2\hat{\bf z}$, and in a given unit cell, corresponding to
the lattice vector ${\bf r}$, the four sites are located at ${\bf r}
-\hat{\bf y}/{2}-\hat{\bf z},~{\bf r} -\hat{\bf y}/{2},~{\bf r}
+\hat{\bf y}/{2}$ and ${\bf r} +\hat{\bf y}/{2}+\hat{\bf z}$.

\subsection{Majorana fermion representation and ground state}

Kitaev mapped the original spin Hamiltonian to a free fermion one
using a Majorana fermion representation of the spin variables. At each
site $j$ he introduced four Majorana fermion operators $\gamma_j^x,~
\gamma_j^y,~\gamma_j^z,~ \eta_j$; different Majorana operators
anticommute, and the square of each of them equals 1. The operators
$\tilde \sigma^\alpha_j = i \gamma_j^\alpha \eta_j$ commute with $D_k
= \gamma_k^x \gamma_k^y \gamma_k^z \eta_k$ for all values of
$\alpha,~j$ and $k$. Moreover, $D_j^2=1$, thus its eigenvalues are $\pm
1$. In the subspace with $D_j=1$, $\tilde\sigma_j^\alpha$ satisfy the
spin-$1/2$ algebra. Thus, in the enlarged space of Majorana fermions
(four-dimensional at each site) the physical states correspond to
$D_j=1$.

In terms of the Majorana operators the Hamiltonian becomes
\begin{align}
\widetilde{H}=\frac{i}{2}\sum\limits_{j,k} J_{\alpha_{jk}}\hat
u_{jk}\eta_j\eta_k,
\end{align}
where $\alpha_{jk}$ is the type of the link between $j$ and $k$, and
$\hat u_{jk}=i\gamma_j^{\alpha_{jk}} \gamma_k^{\alpha_{jk}}$.

$[\hat u_{jk}, \widetilde{H}]=0$ and $[\hat u_{jk}, \hat
  u_{lm}]=0$. In the eigenbasis of $\hat u_{jk}$ the Hamiltonian
becomes
  \begin{align}
\widetilde{H}(\{u_{jk}\})=\frac{i}{2}\sum\limits_{j,k} J_{\alpha_{jk}}
u_{jk}\eta_j\eta_k,
\end{align}
  where $u_{jk}$ is now the eigenvalue of $\hat u_{jk}$. Thus, we have
  mapped the spin model to a system of free fermions in the presence
  of a static $Z_2$ gauge field.

  \subsection{Ground state}

  To find the ground state, we first note that the elements of the
  gauge group are $\prod_j D_j^{n_j}$, where $n_j=0~ \mathrm{or}~
  1$. Under a gauge transformation, $u_{ij} \rightarrow X_i u_{ij}
  X_j$, where $X_i = (1-2n_i)$. The gauge invariant quantities then
  are the Wilson-loop variables $W_l = \prod_{<ij>} u_{ij}$, where
  $<ij>$ are the links belonging to the loop $l$. The elementary
  loops, called the plaquettes, are the smallest loops in the lattice.
  Following a theorem by Lieb\cite{Lie(94)}, it has been shown that the
  ground state corresponds to $W_p = 1$ for all plaquettes
  $p$\cite{Kit(06),ManSur(09)}. To get the physical ground state, we
  first find the lowest energy state in any one of the $\{u_{ij}\}$
  configurations for which $W_p=1$ for all $p$, and then project it to
  $D_j=1$ subspace.

The total Hilbert space is the tensor product of the gauge sector and
the fermion sector. Let $u$ denote a $\{u_{ij}\}$ configuration for
which $W_p=1$ and let $\phi(u)$ be the corresponding lowest energy
fermion wave function. Then the normalized ground state is (assuming
periodic boundary conditions)
\begin{align}
  |GS\rangle &= \frac{1}{\sqrt{2^{N+1}}}\prod_j
  \left(\frac{1+D_j}{2}\right) |u\rangle \otimes |\phi(u)\rangle,
  \label{e-gs-1}
\end{align}

Elements of the gauge group are products of $D_j$ over all possible
subsets $g$ of the lattice sites: $D_g = \prod_{j\in g} D_j$. The the
ground state can be written as follows.
\begin{align}
  |GS\rangle &= \frac{1}{\sqrt{2^{N+1}}}\sum_g D_g |u\rangle \otimes
  \phi(u)\rangle,
  \label{e-gs}
\end{align}

\section{\label{s-tee} Entanglement entropy}

We now calculate entanglement entropy for the above ground state. Our
calculation is a straightforward generalization of Yao and Qi's
computation for the two-dimensional Kitaev model\cite{YaoQi(10)}, from
hereon referred to as YQ.

Entanglement entropy $S$  between two partitions $A$ and $B$ of a system is
defined as the von Neumann entropy of the reduced density matrix of
one of the partitions:
\begin{align}
  S &= -\mathrm{Tr}~\rho_A \ln \rho_A,
  \end{align}
where $\rho_A = \mathrm{Tr}_B~\rho$, with $\mathrm{Tr}_B$ denoting
partial trace with respect to partition $B$, and $\rho =
|GS\rangle\langle GS|$ is the total density matrix. Note that $S$ is
symmetric under the interchange of $A$ and $B$, i.e., we can also
write $S = -\mathrm{Tr}~\rho_B \ln \rho_B$, where
$\rho_B=\mathrm{Tr}_A~\rho$.

Here a comment is in order regarding partial trace for fermions. Since
spatially separated fermion operators do not commute and are therefore
nonlocal, defining a tensor product state between two partitions with
respect to these degrees of freedom is ambiguous. However, in our case
the physical spin degrees of freedom are quadratic in fermion
operators and the latter can be treated as local since the product of
any pair of fermion operators belonging to one partition will commute
with a product of any pair in the other partition. Therefore, we can
perform partial trace without any ambiguity.

We now briefly go through the steps in YQ and show that their
calculation can be readily extended to the hyperhoneycomb
lattice.

\begin{figure}
    \includegraphics[scale=.8]{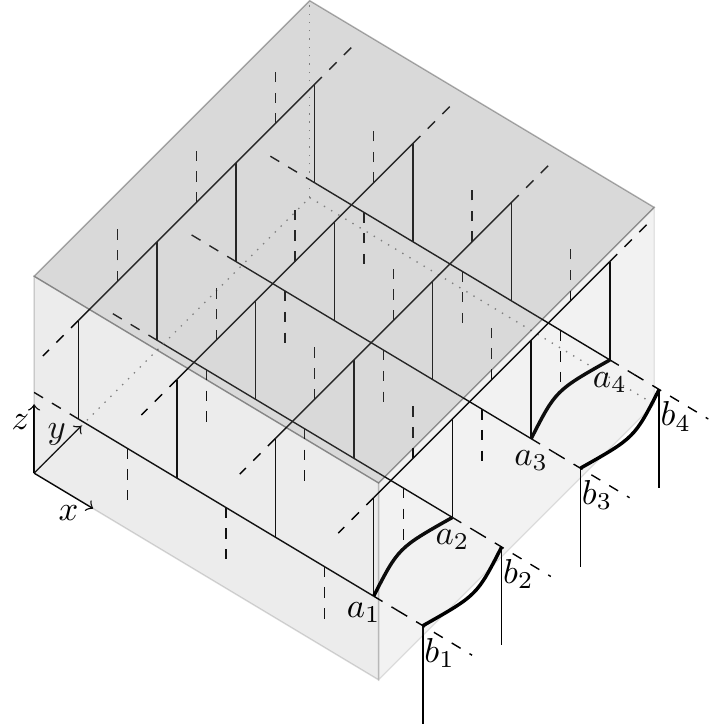}
    \caption{\label{f-sphere-bipart} Bipartition scheme in which region
      $A$ has the topology of a solid sphere. The dashed lines are the
      links on the boundary. $u_{ij}$ variables on the boundary links
      are transformed to $w_{A,n}$ and $w_{B,n}$, which are defined on
      the links $(a_{2n-1}, a_{2n})$ and $(b_{2n-1}, b_{2n})$,
      respectively.}
            \end{figure}

They calculated entanglement entropy using the following replica
method formula\cite{CalCar(04)}:
\begin{align}
S=-\text{Tr}_{A}[\rho_{A}\thinspace \text{log} \thinspace
  \rho_{A}]=\left. -\frac{\partial}{\partial n}
\text{Tr}_{A}[\rho_{A}^{n}]\right|_{n=1}
\label{e-repformula}
\end{align}
To obtain $\rho_A$ we need to do the partial trace over $B$,
$\text{Tr}_B$, and for that we require a set of basis vectors of the
form $|\psi^i\rangle_A \otimes |\chi^i\rangle_B$. But the gauge field
$u_{ij}$ are located at the links and in any partitioning of the
lattice into two regions $A$ and $B$, there will be some links
straddling both $A$ and $B$. To get around this, YQ transformed each
pair of $u_{ij}$ on the shared links into two new variables, one of
them defined on a link lying entirely in region $A$ and the other in
$B$. This is a crucial step in their calculation and is not specific
to two dimensions. In the 3D lattice also the links shared by both
regions can be similarly paired and the corresponding gauge variables
can then be transformed to links lying entirely in either $A$ or $B$
(see Fig. \ref{f-sphere-bipart}). This procedure will be made more
precise when we calculate $S_G$, the contribution to entanglement
entropy from the gauge sector, in the appendix. 

The calculation for the hyperhoneycomb lattice proceeds exactly as in
YQ and we can directly take their following main result (for details
we refer to their paper\cite{YaoQi(10)} and the associated
supplementary material):
\begin{align}
\text{Tr}_{A}[\rho_{A}^n]&= \text{Tr}_{A,G}[\rho_{A,G}^n]
\cdot \text{Tr}_{A,F}[\rho_{A,F}^n],
\label{e-reddm}
\end{align}
where $\rho_{A,F} = \text{Tr}_B|\phi(u)\rangle\langle\phi(u)|$ and
$\rho_{A,G} = \text{Tr}_B|G(u)\rangle\langle G(u)|$ are, respectively,
the reduced density matrix for the Majorana fermion wave function
$|\phi(u)\rangle$ and for the state $|G(u)\rangle =
\left(1/\sqrt{2^{(N-1)}}\right) \sum_{\tilde u} |\tilde u\rangle$ in
the gauge sector. Here $\tilde u$ summation is over all gauge field
configurations gauge equivalent to $u$.

From Eqs. (\ref{e-repformula}) and (\ref{e-reddm}) it immediately
follows that the entanglement entropy $S = S_G + S_F$, where $S_G$ is
the entanglement entropy of the gauge part and $S_F$ that of the
fermionic part. YQ have further shown that $S_F$ has no constant term
independent of the length/area of the boundary, therefore, $S_F$ does not
contribute to TEE.

Calculation of $S_G$ proceeds in exactly the same way as in YQ and
the details are given in the appendix. In our calculation, we also obtain
the dependence of TEE on $b_0$ and $b_1$. Finally, we get
\begin{align}
  S_G &= L \ln 2 - b_0 \ln 2,
  \label{e-entgauge}
  \end{align}
where $2L$ is the number of links on the boundary. Thus $S_G$ depends
only on $b_0$ but not on $b_1$.

\subsection{Topological entanglement entropy}

As discussed in the introduction, the constant term by itself is not a
signature of topological order\cite{GroTur(11)}. Moreover, in the
expression for $S$ in general it is difficult to unambiguously
separate the area term and the constant. However, TEE can still be
extracted using a scheme introduced for 2D systems in
Refs. \onlinecite{KitPre(06)} and \onlinecite{LevWen(06)}. Here we
follow a generalization of this scheme to three
dimensions\cite{CasCha(08)}.

The basic idea is to consider a few different regions of the lattice
and then to take a linear combination of corresponding entanglement
entropies in such a way that all the surface contributions mutually
cancel and the resultant entity is a topological invariant, which can
then be taken as the topological entanglement entropy of the system.

\begin{figure}
    \includegraphics[scale=.8]{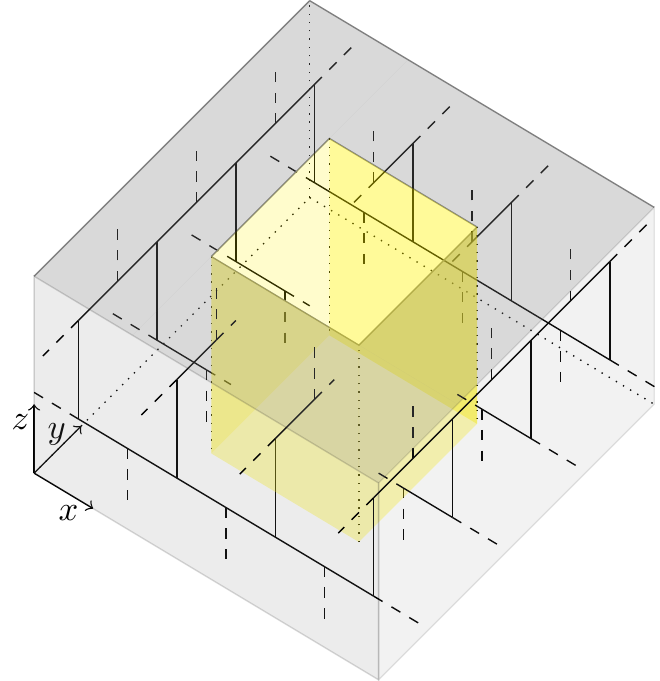}
    \caption{\label{f-torus-bipart} Bipartition scheme in which region
      $A$ is a solid torus. The dashed lines are the links on the
      boundary}
    \end{figure}

\begin{figure}

\subfloat{%
  \includegraphics[clip,width=\columnwidth]{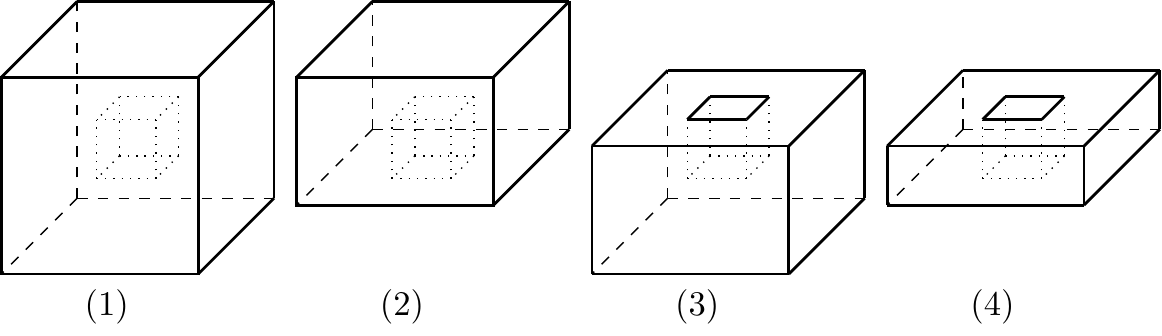}%
}

\subfloat{%
  \includegraphics[clip,width=\columnwidth]{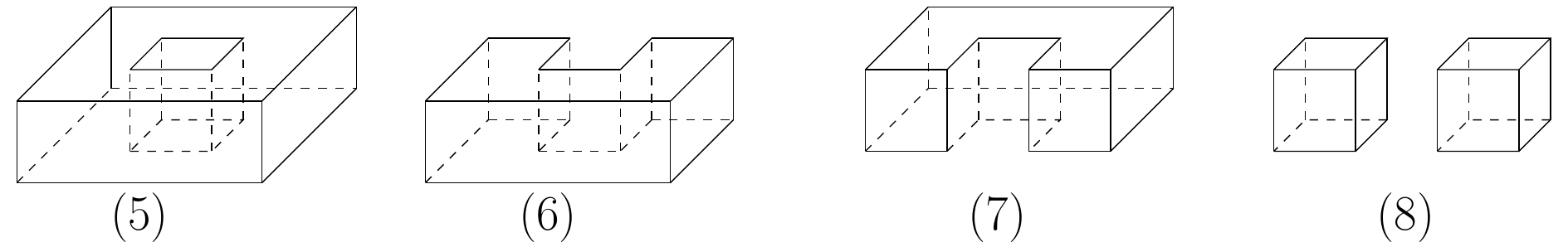}%
}

\caption{\label{f-schemes} Various regions considered for the
      calculation of TEE in the sphere (1-4) and torus (5-8)
      bipartition schemes.} 

\end{figure}

We consider two different bipartitions in which region $A$ is: 1) a
spherical shell, which is nontrivial with respect to closed surfaces,
and 2) a solid torus, which is nontrivial with respect to closed loops
(see Fig. \ref{f-torus-bipart}). In the first case we consider the 
four regions in $A$ shown in Fig. \ref{f-schemes} (1-4). Let
$S_i$ be the entanglement entropy corresponding to the
$i^{\mathrm{th}}$ region. Then using Eq. (\ref{e-entgauge}) we obtain
TEE, $S_{top}^{(1)}$, as
\begin{align}
  S_{top}^{(1)} &= -S_1 + S_2 + S_3 - S_4 = \ln 2. 
\end{align}
In the second case we consider the regions (5-8) shown in
Fig. \ref{f-schemes}, and we get
\begin{align}
  S_{top}^{(2)} &= -S_5 + S_6 + S_7 - S_8 = \ln 2.
\end{align}  
In both the schemes the boundary contributions from various regions
cancel in $S_{top}$ and it is thus invariant under continuous
deformations\cite{KitPre(06),LevWen(06)}.

\section{\label{s-conclusion} Summary and discussion}

We have calculated the topological entanglement entropy for a
three-dimensional hyperhoneycomb lattice generalization of Kitaev's
honeycomb lattice model. We have found that $\gamma_0$, the part of
TEE proportional to $b_0$, is $\ln 2$. The total quantum dimension
$\mathcal{D}$ of this model\cite{ManSur(09)} is $\sqrt{2}$ and
therefore it provides an example of a 3D system in which the relation
$\gamma_0 =\ln \mathcal{D}$ does not hold.

Here $\gamma_0$ is actually $\ln |G|$, where $|G|$ is the order of the
gauge group $G$, in this case $Z_2$. Thus, quite possibly, the
reason $\gamma_0$ is not related to $\mathcal{D}$ in the standard way
for the 3D Kitaev model is that for the latter $\mathcal{D} \neq |G|$.

The low-energy effective Hamiltonian for the Kitaev model in the limit
$J_z \gg J_x, J_y$ is a toric-code-type model defined on the diamond
lattice\cite{ManSur(09),ManSur(14)}. Since TEE for Kitaev model is
independent of the coupling parameters $J_\alpha$, our calculation
provides TEE for the latter model as well.

The effective Hamiltonian also gives a clue as to why TEE for 3D
Kitaev model is different from that for $Z_2$ gauge theory. Toric code
Hamiltonian (in 2D as well as its standard generalization in 3D)
consists of star and plaquette operators. It can then be thought of as
a $Z_2$ gauge theory, with the star operators forming the elements of
the gauge group. However, for the diamond lattice toric code the
operators in the Hamiltonian do not divide into star and plaquette
operators in any obvious manner and such a correspondence with gauge
theory does not exist. Thus we cannot expect the general result for
TEE for gauge theories\cite{GroTur(11)} to hold in this case.

It will be interesting to further explore the general relations among
topological entanglement entropy, gauge group and total quantum
dimension in three dimensions.

\acknowledgements
We thank Saptarshi Mandal for useful discussions. 

\appendix*

\section{Entanglement entropy of the gauge sector}
Our calculation of $S_G$ proceeds as in YQ, and differs from the
latter only in that we additionally obtain the explicit dependence on
$b_0$ and $b_1$.

The full density matrix in the gauge sector is
\begin{align}
  \rho_G &= |G(u)\rangle \langle G(u)| = \frac{1}{2^{(N-1)}}
  \sum_{\tilde u \simeq u} |\tilde u\rangle \langle \tilde u|.
\end{align}
To compute the reduced density matrix $\rho_{G,A}$ we have to carry
out partial trace of $\rho_G$ with respect to the variables in
$B$. But, as pointed out earlier, the variables on the links on the
boundary surface between $A$ and $B$ belong to both the regions. In
YQ, this difficulty is circumvented by the following
procedure.

We can write $|u\rangle = |u_A, u_B, u_p\rangle$, where $u_A$
variables are defined on links entirely in $A$, $u_B$ on links
entirely in $B$, and $u_p$ on links on the boundary and shared by both
$A$ and $B$. Assuming that the number of boundary links is even, and
denoting it by $2L$, we label the corresponding link variables $u_p$ as
$u_{a_1, b_1}, u_{a_2, b_2}, \cdots u_{a_{2L}, b_{2L}}$, where the
sites labeled $a_j$ are in $A$ and those labeled $b_j$ are in $B$. In
terms of Majorana variables, $\hat u_{a_j,b_j} = i
\gamma_{a_j}^{\alpha_j} \gamma_{b_j}^{\alpha_j}$, where $\alpha_j$ is
the link-type of $(a_j, b_j)$. Now define new variables $\hat w_{A,n}
= i\gamma_{a_{(2n-1)}}^{\alpha_{(2n-1)}}
\gamma_{a_{2n}}^{\alpha_{2n}}$ and $\hat w_{B,n} =
i\gamma_{b_{(2n-1)}}^{\alpha_{(2n-1)}}
\gamma_{b_{2n}}^{\alpha_{2n}}$. $\hat w_{A,n}$ is defined on the link
$(a_{2n-1},a_{2n})$, which lies entirely in $A$; similarly, $\hat
w_{B,n}$ is defined on $(b_{2n-1}, b_{2n})$, which lies entirely in
$B$ (see Fig. \ref{f-sphere-bipart}).

Since $\{u_{ij}\}$ is any gauge-field configuration for which $W_p=1$ for
all plaquettes, we can choose $u_{a_j,b_j}= 1$ for all the boundary
links. Then, it is easy to verify that
\begin{align}
  |u_p \rangle = \frac{1}{\sqrt{2^L}} \sum_{w_{A}=w_{B}=\pm 1} | w_{A},
  w_{B}\rangle,
  \label{e-utow}
  \end{align}
where $w_A$ and $w_B$ denote the set of eigenvalues of $\hat w_{A,n}$
and $\hat w_{B,n}$, respectively. Thus,
\begin{widetext}
\begin{align}
  |G(u) \rangle = \frac{1}{\sqrt{2^{N+L+1}}} \sum_g \sum_{w_A=w_B} D_g
  |u_A, w_A; u_B, w_B\rangle.
\end{align}
Writing $D_g = X_{g_A}\cdot X_{g_B}$, where $g_A$ is the set of sites
in $g$ belonging to $A$ and $X_{g_A}= \prod_{j\in g_A}D_j$. $X_{g_B}$
is similarly defined. Then,
\begin{align}
  \rho_{G,A} = \mathrm{Tr}_B \rho_{G} &= \frac{1}{2^{N+L+1}}
  \sum_{g,g'}\sum_{w,w'} X_{g_A} |u_A,w\rangle \langle u_A,w'|X_{g'_A}^\dagger  
\sum_{u'_B,w''} \langle u'_B, w''| X_{g_B} |u_B,w\rangle \langle
u_B,w'|X_{g'_B}^\dagger|u'_B, w'' \rangle \nonumber \\
  \rho_{G,A} &= \frac{1}{2^{N+L+1}}
  \sum_{g,g'}\sum_{w,w'} X_{g_A} |u_A,w\rangle \langle u_A,w'|X_{g'_A}^\dagger  
\langle u_B,w'|X_{g'_B}^\dagger X_{g_B} |u_B,w\rangle
\end{align}
\end{widetext}
For $\langle u_B,w'|X_{g'_B}^\dagger X_{g_B} |u_B,w\rangle$ to be
nonzero, $w=w'$. Further conditions for its nonvanishing depend on the
topology of region $B$. Let $g_B^{(n)}$, $n=1,\cdots, n_{B}$, denote
the sites in $g_B$ belonging to the connected component $B_n$ of
$B$. Here $n_B$ is the number of connected components of $B$. Then the
nonvanishing condition becomes: for each $n$, either ${g'_B}^{(n)} =
g_B^{(n)}$, for which $X_{{g'_B}^{(n)}}^\dagger X_{g_B^{(n)}} =1$, or
${g'_B}^{(n)} = B_n-{g_B}^{(n)}$, in which case
$X_{{g'_B}^{(n)}}^\dagger X_{{g_B}^{(n)}} =X_{B_n}$ (here
$X_{B_n}\equiv X_{g=B_n}$). In both the cases $\langle
u_B,w'|X_{g'_B}^\dagger X_{g_B} |u_B,w\rangle=1$. Let $N_A$ and $N_B$
be the number of sites in $A$ and $B$, respectively (with
$N_A+N_B=N$). Then,
\begin{align}
  \rho_{G,A} &= \frac{2^{n_{B}}}{2^{N_A+L+1}}
  \sum_{g_A,g'_A,w} X_{g_A} |u_A,w\rangle \langle u_A,w|X_{g'_A}^\dagger.
\end{align}
Next we calculate $\rho_{G,A}^2$ and show that it is proportional to
$\rho_{G,A}$.
\begin{widetext}
\begin{align}
\rho_{G,A}^2 &= \left(\frac{2^{n_{B}}}{2^{N_A+L+1}}\right)^2
  \sum_{\substack{g_A,g'_A,w \\ \tilde g_A,\tilde g'_A,w'}} X_{g_A}
  |u_A,w\rangle \langle u_A,w|X_{g'_A}^\dagger X_{\tilde g_A}
  |u_A,w'\rangle \langle u_A,w'|X_{\tilde g'_A}^\dagger.
\end{align}  
As before, $\langle u_A,w|X_{g'_A}^\dagger X_{\tilde g_A}
|u_A,w'\rangle$ is nonzero only when $w=w'$ and, for each connected
component $A_n$ in $A$, either ${g'_A}^{(n)} = g_A^{(n)}$, or
${g'_A}^{(n)} = A_n-{g_A}^{(n)}$,  (here ${g_A}^{(n)}$ denotes the
sites in $g_A$ belonging to $A_n$). Then
\begin{align}
\rho_{G,A}^2 &= \left(\frac{2^{n_{B}}}{2^{N_A+L+1}}\right)^2 \times 2^{N_A+n_A}
  \sum_{g_A,g'_A,w} X_{g_A} |u_A,w\rangle  \langle u_A,w|X_{g'_A}^\dagger,
\end{align}  
\end{widetext}
where $n_A$ is the number of connected components in $A$. Thus,
\begin{align}
  \rho_{G,A}^2 &= 2^{n_A+n_B-L-1} \rho_{G,A}.
\end{align}
\begin{sloppypar}
From the properties of density matrix it then immediately follows that
the entanglement entropy ${S_G = L \ln 2 - (n_A+n_B-1) \ln 2}$.
But $n_A+n_B-1 = b_0$, the number of connected
components (zeroth Betti number) of the boundary surface between $A$
and $B$, and we have \end{sloppypar}
\begin{align}
  S_G &= L \ln 2 - b_0 \ln 2.
  \end{align}

\bibliography{reference}

\begin{thebibliography}{25}%
\makeatletter
\providecommand \@ifxundefined [1]{%
 \@ifx{#1\undefined}
}%
\providecommand \@ifnum [1]{%
 \ifnum #1\expandafter \@firstoftwo
 \else \expandafter \@secondoftwo
 \fi
}%
\providecommand \@ifx [1]{%
 \ifx #1\expandafter \@firstoftwo
 \else \expandafter \@secondoftwo
 \fi
}%
\providecommand \natexlab [1]{#1}%
\providecommand \enquote  [1]{``#1''}%
\providecommand \bibnamefont  [1]{#1}%
\providecommand \bibfnamefont [1]{#1}%
\providecommand \citenamefont [1]{#1}%
\providecommand \href@noop [0]{\@secondoftwo}%
\providecommand \href [0]{\begingroup \@sanitize@url \@href}%
\providecommand \@href[1]{\@@startlink{#1}\@@href}%
\providecommand \@@href[1]{\endgroup#1\@@endlink}%
\providecommand \@sanitize@url [0]{\catcode `\\12\catcode `\$12\catcode
  `\&12\catcode `\#12\catcode `\^12\catcode `\_12\catcode `\%12\relax}%
\providecommand \@@startlink[1]{}%
\providecommand \@@endlink[0]{}%
\providecommand \url  [0]{\begingroup\@sanitize@url \@url }%
\providecommand \@url [1]{\endgroup\@href {#1}{\urlprefix }}%
\providecommand \urlprefix  [0]{URL }%
\providecommand \Eprint [0]{\href }%
\providecommand \doibase [0]{http://dx.doi.org/}%
\providecommand \selectlanguage [0]{\@gobble}%
\providecommand \bibinfo  [0]{\@secondoftwo}%
\providecommand \bibfield  [0]{\@secondoftwo}%
\providecommand \translation [1]{[#1]}%
\providecommand \BibitemOpen [0]{}%
\providecommand \bibitemStop [0]{}%
\providecommand \bibitemNoStop [0]{.\EOS\space}%
\providecommand \EOS [0]{\spacefactor3000\relax}%
\providecommand \BibitemShut  [1]{\csname bibitem#1\endcsname}%
\let\auto@bib@innerbib\@empty
\bibitem [{\citenamefont {Srednicki}(1993)}]{Sre(93)}%
  \BibitemOpen
  \bibfield  {author} {\bibinfo {author} {\bibfnamefont {M.}~\bibnamefont
  {Srednicki}},\ }\href {\doibase 10.1103/PhysRevLett.71.666} {\bibfield
  {journal} {\bibinfo  {journal} {Phys. Rev. Lett.}\ }\textbf {\bibinfo
  {volume} {71}},\ \bibinfo {pages} {666} (\bibinfo {year} {1993})}\BibitemShut
  {NoStop}%
\bibitem [{\citenamefont {Kitaev}\ and\ \citenamefont
  {Preskill}(2006)}]{KitPre(06)}%
  \BibitemOpen
  \bibfield  {author} {\bibinfo {author} {\bibfnamefont {A.}~\bibnamefont
  {Kitaev}}\ and\ \bibinfo {author} {\bibfnamefont {J.}~\bibnamefont
  {Preskill}},\ }\href {\doibase 10.1103/PhysRevLett.96.110404} {\bibfield
  {journal} {\bibinfo  {journal} {Phys. Rev. Lett.}\ }\textbf {\bibinfo
  {volume} {96}},\ \bibinfo {pages} {110404} (\bibinfo {year}
  {2006})}\BibitemShut {NoStop}%
\bibitem [{\citenamefont {Levin}\ and\ \citenamefont {Wen}(2006)}]{LevWen(06)}%
  \BibitemOpen
  \bibfield  {author} {\bibinfo {author} {\bibfnamefont {M.}~\bibnamefont
  {Levin}}\ and\ \bibinfo {author} {\bibfnamefont {X.-G.}\ \bibnamefont
  {Wen}},\ }\href {\doibase 10.1103/PhysRevLett.96.110405} {\bibfield
  {journal} {\bibinfo  {journal} {Phys. Rev. Lett.}\ }\textbf {\bibinfo
  {volume} {96}},\ \bibinfo {pages} {110405} (\bibinfo {year}
  {2006})}\BibitemShut {NoStop}%
\bibitem [{\citenamefont {Kitaev}(2003)}]{Kit(03)}%
  \BibitemOpen
  \bibfield  {author} {\bibinfo {author} {\bibfnamefont {A.}~\bibnamefont
  {Kitaev}},\ }\href {\doibase http://dx.doi.org/10.1016/S0003-4916(02)00018-0}
  {\bibfield  {journal} {\bibinfo  {journal} {Annals of Physics}\ }\textbf
  {\bibinfo {volume} {303}},\ \bibinfo {pages} {2 } (\bibinfo {year}
  {2003})}\BibitemShut {NoStop}%
\bibitem [{\citenamefont {Hamma}\ \emph {et~al.}(2005)\citenamefont {Hamma},
  \citenamefont {Ionicioiu},\ and\ \citenamefont {Zanardi}}]{HamIon(05)}%
  \BibitemOpen
  \bibfield  {author} {\bibinfo {author} {\bibfnamefont {A.}~\bibnamefont
  {Hamma}}, \bibinfo {author} {\bibfnamefont {R.}~\bibnamefont {Ionicioiu}}, \
  and\ \bibinfo {author} {\bibfnamefont {P.}~\bibnamefont {Zanardi}},\ }\href
  {\doibase http://dx.doi.org/10.1016/j.physleta.2005.01.060} {\bibfield
  {journal} {\bibinfo  {journal} {Physics Letters A}\ }\textbf {\bibinfo
  {volume} {337}},\ \bibinfo {pages} {22 } (\bibinfo {year}
  {2005})}\BibitemShut {NoStop}%
\bibitem [{\citenamefont {Kitaev}(2006)}]{Kit(06)}%
  \BibitemOpen
  \bibfield  {author} {\bibinfo {author} {\bibfnamefont {A.}~\bibnamefont
  {Kitaev}},\ }\href {\doibase http://dx.doi.org/10.1016/j.aop.2005.10.005}
  {\bibfield  {journal} {\bibinfo  {journal} {Annals of Physics}\ }\textbf
  {\bibinfo {volume} {321}},\ \bibinfo {pages} {2 } (\bibinfo {year}
  {2006})}\BibitemShut {NoStop}%
\bibitem [{\citenamefont {Yao}\ and\ \citenamefont {Qi}(2010)}]{YaoQi(10)}%
  \BibitemOpen
  \bibfield  {author} {\bibinfo {author} {\bibfnamefont {H.}~\bibnamefont
  {Yao}}\ and\ \bibinfo {author} {\bibfnamefont {X.-L.}\ \bibnamefont {Qi}},\
  }\href {\doibase 10.1103/PhysRevLett.105.080501} {\bibfield  {journal}
  {\bibinfo  {journal} {Phys. Rev. Lett.}\ }\textbf {\bibinfo {volume} {105}},\
  \bibinfo {pages} {080501} (\bibinfo {year} {2010})}\BibitemShut {NoStop}%
\bibitem [{\citenamefont {Grover}\ \emph {et~al.}(2011)\citenamefont {Grover},
  \citenamefont {Turner},\ and\ \citenamefont {Vishwanath}}]{GroTur(11)}%
  \BibitemOpen
  \bibfield  {author} {\bibinfo {author} {\bibfnamefont {T.}~\bibnamefont
  {Grover}}, \bibinfo {author} {\bibfnamefont {A.~M.}\ \bibnamefont {Turner}},
  \ and\ \bibinfo {author} {\bibfnamefont {A.}~\bibnamefont {Vishwanath}},\
  }\href {\doibase 10.1103/PhysRevB.84.195120} {\bibfield  {journal} {\bibinfo
  {journal} {Phys. Rev. B}\ }\textbf {\bibinfo {volume} {84}},\ \bibinfo
  {pages} {195120} (\bibinfo {year} {2011})}\BibitemShut {NoStop}%
\bibitem [{\citenamefont {Castelnovo}\ and\ \citenamefont
  {Chamon}(2008)}]{CasCha(08)}%
  \BibitemOpen
  \bibfield  {author} {\bibinfo {author} {\bibfnamefont {C.}~\bibnamefont
  {Castelnovo}}\ and\ \bibinfo {author} {\bibfnamefont {C.}~\bibnamefont
  {Chamon}},\ }\href {\doibase 10.1103/PhysRevB.78.155120} {\bibfield
  {journal} {\bibinfo  {journal} {Phys. Rev. B}\ }\textbf {\bibinfo {volume}
  {78}},\ \bibinfo {pages} {155120} (\bibinfo {year} {2008})}\BibitemShut
  {NoStop}%
\bibitem [{\citenamefont {Iblisdir}\ \emph {et~al.}(2009)\citenamefont
  {Iblisdir}, \citenamefont {P\'erez-Garc\'{\i}a}, \citenamefont {Aguado},\
  and\ \citenamefont {Pachos}}]{IblPer(09)}%
  \BibitemOpen
  \bibfield  {author} {\bibinfo {author} {\bibfnamefont {S.}~\bibnamefont
  {Iblisdir}}, \bibinfo {author} {\bibfnamefont {D.}~\bibnamefont
  {P\'erez-Garc\'{\i}a}}, \bibinfo {author} {\bibfnamefont {M.}~\bibnamefont
  {Aguado}}, \ and\ \bibinfo {author} {\bibfnamefont {J.}~\bibnamefont
  {Pachos}},\ }\href {\doibase 10.1103/PhysRevB.79.134303} {\bibfield
  {journal} {\bibinfo  {journal} {Phys. Rev. B}\ }\textbf {\bibinfo {volume}
  {79}},\ \bibinfo {pages} {134303} (\bibinfo {year} {2009})}\BibitemShut
  {NoStop}%
\bibitem [{\citenamefont {Iblisdir}\ \emph {et~al.}(2010)\citenamefont
  {Iblisdir}, \citenamefont {Pérez-García}, \citenamefont {Aguado},\ and\
  \citenamefont {Pachos}}]{IblPer(10)}%
  \BibitemOpen
  \bibfield  {author} {\bibinfo {author} {\bibfnamefont {S.}~\bibnamefont
  {Iblisdir}}, \bibinfo {author} {\bibfnamefont {D.}~\bibnamefont
  {Pérez-García}}, \bibinfo {author} {\bibfnamefont {M.}~\bibnamefont
  {Aguado}}, \ and\ \bibinfo {author} {\bibfnamefont {J.}~\bibnamefont
  {Pachos}},\ }\href {\doibase https://doi.org/10.1016/j.nuclphysb.2009.11.009}
  {\bibfield  {journal} {\bibinfo  {journal} {Nuclear Physics B}\ }\textbf
  {\bibinfo {volume} {829}},\ \bibinfo {pages} {401 } (\bibinfo {year}
  {2010})}\BibitemShut {NoStop}%
\bibitem [{\citenamefont {Walker}\ and\ \citenamefont
  {Wang}(2012)}]{WalWan(12)}%
  \BibitemOpen
  \bibfield  {author} {\bibinfo {author} {\bibfnamefont {K.}~\bibnamefont
  {Walker}}\ and\ \bibinfo {author} {\bibfnamefont {Z.}~\bibnamefont {Wang}},\
  }\href {\doibase 10.1007/s11467-011-0194-z} {\bibfield  {journal} {\bibinfo
  {journal} {Frontiers of Physics}\ }\textbf {\bibinfo {volume} {7}},\ \bibinfo
  {pages} {150} (\bibinfo {year} {2012})}\BibitemShut {NoStop}%
\bibitem [{\citenamefont {von Keyserlingk}\ \emph {et~al.}(2013)\citenamefont
  {von Keyserlingk}, \citenamefont {Burnell},\ and\ \citenamefont
  {Simon}}]{KeyBur(13)}%
  \BibitemOpen
  \bibfield  {author} {\bibinfo {author} {\bibfnamefont {C.~W.}\ \bibnamefont
  {von Keyserlingk}}, \bibinfo {author} {\bibfnamefont {F.~J.}\ \bibnamefont
  {Burnell}}, \ and\ \bibinfo {author} {\bibfnamefont {S.~H.}\ \bibnamefont
  {Simon}},\ }\href {https://link.aps.org/doi/10.1103/PhysRevB.87.045107}
  {\bibfield  {journal} {\bibinfo  {journal} {Phys. Rev. B}\ }\textbf {\bibinfo
  {volume} {87}},\ \bibinfo {pages} {045107} (\bibinfo {year}
  {2013})}\BibitemShut {NoStop}%
\bibitem [{\citenamefont {Bullivant}\ and\ \citenamefont
  {Pachos}(2016)}]{BulPac(16)}%
  \BibitemOpen
  \bibfield  {author} {\bibinfo {author} {\bibfnamefont {A.}~\bibnamefont
  {Bullivant}}\ and\ \bibinfo {author} {\bibfnamefont {J.~K.}\ \bibnamefont
  {Pachos}},\ }\href {\doibase 10.1103/PhysRevB.93.125111} {\bibfield
  {journal} {\bibinfo  {journal} {Phys. Rev. B}\ }\textbf {\bibinfo {volume}
  {93}},\ \bibinfo {pages} {125111} (\bibinfo {year} {2016})}\BibitemShut
  {NoStop}%
\bibitem [{\citenamefont {Mondragon-Shem}\ and\ \citenamefont
  {Hughes}(2014)}]{MonHug(14)}%
  \BibitemOpen
  \bibfield  {author} {\bibinfo {author} {\bibfnamefont {I.}~\bibnamefont
  {Mondragon-Shem}}\ and\ \bibinfo {author} {\bibfnamefont {T.~L.}\
  \bibnamefont {Hughes}},\ }\href
  {http://stacks.iop.org/1742-5468/2014/i=10/a=P10022} {\bibfield  {journal}
  {\bibinfo  {journal} {Journal of Statistical Mechanics: Theory and
  Experiment}\ }\textbf {\bibinfo {volume} {2014}},\ \bibinfo {pages} {P10022}
  (\bibinfo {year} {2014})}\BibitemShut {NoStop}%
\bibitem [{\citenamefont {Ryu}(2009)}]{Ryu(09)}%
  \BibitemOpen
  \bibfield  {author} {\bibinfo {author} {\bibfnamefont {S.}~\bibnamefont
  {Ryu}},\ }\href {\doibase 10.1103/PhysRevB.79.075124} {\bibfield  {journal}
  {\bibinfo  {journal} {Phys. Rev. B}\ }\textbf {\bibinfo {volume} {79}},\
  \bibinfo {pages} {075124} (\bibinfo {year} {2009})}\BibitemShut {NoStop}%
\bibitem [{\citenamefont {Mandal}\ and\ \citenamefont
  {Surendran}(2009)}]{ManSur(09)}%
  \BibitemOpen
  \bibfield  {author} {\bibinfo {author} {\bibfnamefont {S.}~\bibnamefont
  {Mandal}}\ and\ \bibinfo {author} {\bibfnamefont {N.}~\bibnamefont
  {Surendran}},\ }\href {\doibase 10.1103/PhysRevB.79.024426} {\bibfield
  {journal} {\bibinfo  {journal} {Phys. Rev. B}\ }\textbf {\bibinfo {volume}
  {79}},\ \bibinfo {pages} {024426} (\bibinfo {year} {2009})}\BibitemShut
  {NoStop}%
\bibitem [{\citenamefont {Kimchi}\ \emph {et~al.}(2014)\citenamefont {Kimchi},
  \citenamefont {Analytis},\ and\ \citenamefont {Vishwanath}}]{KimAna(14)}%
  \BibitemOpen
  \bibfield  {author} {\bibinfo {author} {\bibfnamefont {I.}~\bibnamefont
  {Kimchi}}, \bibinfo {author} {\bibfnamefont {J.~G.}\ \bibnamefont
  {Analytis}}, \ and\ \bibinfo {author} {\bibfnamefont {A.}~\bibnamefont
  {Vishwanath}},\ }\href {\doibase 10.1103/PhysRevB.90.205126} {\bibfield
  {journal} {\bibinfo  {journal} {Phys. Rev. B}\ }\textbf {\bibinfo {volume}
  {90}},\ \bibinfo {pages} {205126} (\bibinfo {year} {2014})}\BibitemShut
  {NoStop}%
\bibitem [{\citenamefont {O'Brien}\ \emph {et~al.}(2016)\citenamefont
  {O'Brien}, \citenamefont {Hermanns},\ and\ \citenamefont
  {Trebst}}]{ObrHer(16)}%
  \BibitemOpen
  \bibfield  {author} {\bibinfo {author} {\bibfnamefont {K.}~\bibnamefont
  {O'Brien}}, \bibinfo {author} {\bibfnamefont {M.}~\bibnamefont {Hermanns}}, \
  and\ \bibinfo {author} {\bibfnamefont {S.}~\bibnamefont {Trebst}},\ }\href
  {\doibase 10.1103/PhysRevB.93.085101} {\bibfield  {journal} {\bibinfo
  {journal} {Phys. Rev. B}\ }\textbf {\bibinfo {volume} {93}},\ \bibinfo
  {pages} {085101} (\bibinfo {year} {2016})}\BibitemShut {NoStop}%
\bibitem [{\citenamefont {Matern}\ and\ \citenamefont
  {Hermanns}(2017)}]{MatHer(17)}%
  \BibitemOpen
  \bibfield  {author} {\bibinfo {author} {\bibfnamefont {S.}~\bibnamefont
  {Matern}}\ and\ \bibinfo {author} {\bibfnamefont {M.}~\bibnamefont
  {Hermanns}},\ }\href@noop {} {\bibfield  {journal} {\bibinfo  {journal}
  {arXiv:1712.07715}\ } (\bibinfo {year} {2017})}\BibitemShut {NoStop}%
\bibitem [{\citenamefont {Winter}\ \emph {et~al.}(2017)\citenamefont {Winter},
  \citenamefont {Tsirlin}, \citenamefont {Daghofer}, \citenamefont {van~den
  Brink}, \citenamefont {Singh}, \citenamefont {Gegenwart},\ and\ \citenamefont
  {Valenti}}]{WinTsi(17)}%
  \BibitemOpen
  \bibfield  {author} {\bibinfo {author} {\bibfnamefont {S.~M.}\ \bibnamefont
  {Winter}}, \bibinfo {author} {\bibfnamefont {A.~A.}\ \bibnamefont {Tsirlin}},
  \bibinfo {author} {\bibfnamefont {M.}~\bibnamefont {Daghofer}}, \bibinfo
  {author} {\bibfnamefont {J.}~\bibnamefont {van~den Brink}}, \bibinfo {author}
  {\bibfnamefont {Y.}~\bibnamefont {Singh}}, \bibinfo {author} {\bibfnamefont
  {P.}~\bibnamefont {Gegenwart}}, \ and\ \bibinfo {author} {\bibfnamefont
  {R.}~\bibnamefont {Valenti}},\ }\href
  {http://stacks.iop.org/0953-8984/29/i=49/a=493002} {\bibfield  {journal}
  {\bibinfo  {journal} {Journal of Physics: Condensed Matter}\ }\textbf
  {\bibinfo {volume} {29}},\ \bibinfo {pages} {493002} (\bibinfo {year}
  {2017})}\BibitemShut {NoStop}%
\bibitem [{\citenamefont {Jackeli}\ and\ \citenamefont
  {Khaliullin}(2009)}]{JacKha(09)}%
  \BibitemOpen
  \bibfield  {author} {\bibinfo {author} {\bibfnamefont {G.}~\bibnamefont
  {Jackeli}}\ and\ \bibinfo {author} {\bibfnamefont {G.}~\bibnamefont
  {Khaliullin}},\ }\href {\doibase 10.1103/PhysRevLett.102.017205} {\bibfield
  {journal} {\bibinfo  {journal} {Phys. Rev. Lett.}\ }\textbf {\bibinfo
  {volume} {102}},\ \bibinfo {pages} {017205} (\bibinfo {year}
  {2009})}\BibitemShut {NoStop}%
\bibitem [{\citenamefont {Lieb}(1994)}]{Lie(94)}%
  \BibitemOpen
  \bibfield  {author} {\bibinfo {author} {\bibfnamefont {E.~H.}\ \bibnamefont
  {Lieb}},\ }\href {\doibase 10.1103/PhysRevLett.73.2158} {\bibfield  {journal}
  {\bibinfo  {journal} {Phys. Rev. Lett.}\ }\textbf {\bibinfo {volume} {73}},\
  \bibinfo {pages} {2158} (\bibinfo {year} {1994})}\BibitemShut {NoStop}%
\bibitem [{\citenamefont {Calabrese}\ and\ \citenamefont
  {Cardy}(2004)}]{CalCar(04)}%
  \BibitemOpen
  \bibfield  {author} {\bibinfo {author} {\bibfnamefont {P.}~\bibnamefont
  {Calabrese}}\ and\ \bibinfo {author} {\bibfnamefont {J.}~\bibnamefont
  {Cardy}},\ }\href {http://stacks.iop.org/1742-5468/2004/i=06/a=P06002}
  {\bibfield  {journal} {\bibinfo  {journal} {Journal of Statistical Mechanics:
  Theory and Experiment}\ }\textbf {\bibinfo {volume} {2004}},\ \bibinfo
  {pages} {P06002} (\bibinfo {year} {2004})}\BibitemShut {NoStop}%
\bibitem [{\citenamefont {Mandal}\ and\ \citenamefont
  {Surendran}(2014)}]{ManSur(14)}%
  \BibitemOpen
  \bibfield  {author} {\bibinfo {author} {\bibfnamefont {S.}~\bibnamefont
  {Mandal}}\ and\ \bibinfo {author} {\bibfnamefont {N.}~\bibnamefont
  {Surendran}},\ }\href {\doibase 10.1103/PhysRevB.90.104424} {\bibfield
  {journal} {\bibinfo  {journal} {Phys. Rev. B}\ }\textbf {\bibinfo {volume}
  {90}},\ \bibinfo {pages} {104424} (\bibinfo {year} {2014})}\BibitemShut
  {NoStop}%
\end{thebibliography}%

\end{document}